\documentclass[pra,twocolumn,showpacs,superscriptaddress,groupedaddress,floatfix]{revtex4-1}
\usepackage{amsmath,amssymb}
\usepackage{newtxtext,newtxmath}
\usepackage{graphicx}

\newcommand{\ingaas}{{$\text{In}_\text{0.47}\text{Ga}_\text{0.53}\text{As}$}}
\newcommand{\um}{{um}}
\newcommand{\us}{{us}}

\begin{document}

\title{Excess Loss in Homodyne Detection\\
Originating from Distributed Photocarrier Generation in Photodiodes}
\author{Takahiro Serikawa and Akira Furusawa}
\email[]{akiraf@ap.t.u-tokyo.ac.jp}
\affiliation{Department of Applied Physics, School of Engineering, The University of Tokyo, 7-3-1 Hongo, Bunkyo-ku, Tokyo 113-8656, Japan.}

\date{\today}

\begin{abstract}
  The distributed absorption of photons in photodiodes induces an excess noise in continuous-wave photodetection above the transit-time roll-off frequency. We show that it can be treated as a frequency-dependent excess optical loss in homodyne detection. This places a limit on the bandwidth of high-accuracy homodyne detection, even if an ideal photodetector circuit is available. We experimentally verify the excess loss in two ways; a comparison of signal gain and shot-noise gain of one-port homodyne detection, and a balanced homodyne detection of squeezed light at 500\,MHz sideband. These results agree with an analytic expression we develop, where the randomness of the photoabsorption is directly modeled by an intrusion of vacuum field. At 500\,MHz, we estimate the excess loss at 14\% for a Si-PIN photodiode with 860\,nm incident light, while the numerical simulation predicts much smaller excess loss in InGaAs photodiodes with 1550\,nm light. 
\end{abstract}

\maketitle

\section{Introduction}
Optical homodyne detection \cite{Yuen:83,Schumaker:84} is a key technology in photonic quantum communication and computation such as continuous-variable (CV) quantum key distribution \cite{PhysRevLett.67.661} or CV measurement-based computation \cite{PhysRevA.64.012310,PhysRevLett.90.117901,PhysRevLett.97.110501}, since it offers high-speed, high-precision quantum measurement. For these applications, optical loss in the interferometer or the photodiode must be kept low to avoid mixing the target quantum observable with vacuum noise. Even a few percents of loss can be a fatal issue in the struggle to exceed the quantum error correction threshold \cite{PhysRevLett.112.120504}. Also, any other noises in the detection process degrade the signal and can cause errors in the quantum protocols. For the future practical applications of homodyne detection, it is especially challenging but important to realize high-speed detectors while avoiding losses and noises. Recent development of highly-efficient, high-speed, low-noise homodyne detectors has realized a wideband detector with more than 100\,MHz bandwidth \cite{KUMAR20125259,0256-307X-30-11-114209} or a narrowband detector whose center frequency is as high as 500\,MHz \cite{doi:10.1063/1.5029859}. In principle, the bandwidth of homodyne detectors is not limited by the cut-off of the photodiodes or detector circuits because it can be canceled by post-processing equalizers. The signal-to-noise ratio is the true limiting factor, while it can be possibly improved by cryogenic amplifiers that have quite low noise level. Here a question arises; what is the fundamental limit of the bandwidth of quantum homodyne detection?

Sun \textit{et al.} pointed out \cite{PhysRevLett.113.203901} that distributed absorption or randomness in photocarrier transport in photodiodes cause excess photocurrent noise. This excess noise becomes an intrinsic limit of the signal-to-noise ratio of photodetection in the high-frequency domain. They showed that carrier scattering is the major factor of the excess noise for InGaAs uni-traveling-carrier photodiodes over 10\,GHz, and it increases timing jitter of the detection of ultrashort optical pulses.

This mechanism also takes place in continuous-wave (CW) photodetection, leading to a broadband noise limit in homodyne detection. In this paper, we introduce a simple model using a series beamsplitter picture and show that the randomness of the distributed absorption is described by a frequency-dependent optical loss. We experimentally examine the loss spectrum of Si-PIN photodiodes by two ways: shot-noise spectrum measurement and squeezed light measurement. We show that the distributed absorption is dominant in the excess noise of Si-PIN photodiodes with a thick absorption layer and that the excess loss is not ignorable above 100\,MHz region, which will become an urgent problem for the forthcoming broadband optical quantum information processing.

\section{Model}
\subsection{Noise in Homodyne detection}
Optical homodyne detection \cite{Yuen:83,Schumaker:84} can be considered as demodulation of signal light field with a local oscillator (LO) at the carrier light frequency. In an ideal situation, the photocurrent signal $I$ is proportional to the quadrature amplitude of the signal field and homodyne detection is a projection measurement of a quadrature operator $\hat{x}$. When the signal light is a vacuum, the random arrival of LO photons, i.e. shot-noise, is interpreted as vacuum fluctuation of $\hat{x}$. $I$ is normalized by the shot-noise amplitude $\sqrt{\langle I_\text{shot}^2 \rangle}$, which is usually done in frequency-domain as $x(\omega) = I(\omega) \big/ \sqrt{\langle I_\text{shot}(\omega)^2 \rangle}$ for the compensation of the frequency-dependent gain of the detector. Here a convenient way of treating extra Gaussian noise in the detection process such as the electric noise of the detector is to include it in the shot-noise, and the normalization procedure in homodyne detection reinterpret the extra noise as a vacuum noise, i.e. an optical loss \cite{PhysRevA.75.035802}. The excess noise originating from the distributed photocarrier generation or random carrier transport is also to be regarded as an optical loss, since the excess noise is Gaussian due to the central limit theorem for the average of numerous photons in the LO.

\subsection{Effect of Distributed Absorption on Homodyne Detection}
\begin{figure}[tb]
    \centering
    \includegraphics{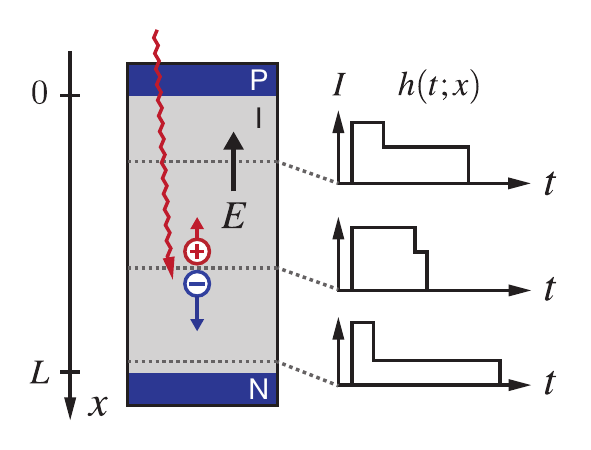}
    \caption{Illustration of distributed photocarrier generation in a PIN photodiode. Position-dependent photocurrent impulse responses $h(t; x)$ are shown at different absorption positions.}
    \label{fig:jitter_schem}
\end{figure}

We consider a photodetection with a simple PIN-photodiode. The input light photons are transformed to electron-hole pairs in the intrinsic region at random penetration depth, which has exponential decay distribution $f(x) = \alpha e^{-\alpha x}$, where $\alpha$ is the absorption coefficient. The carrier transport of each photocarrier pair generates photocurrent at the electrode, which has a position-dependent impulse response function $h(t; x)$ (Fig. \ref{fig:jitter_schem}). Here $h(t; x)$ is assumed to be a deterministic function; the randomness in the transport such as recombination or scattering is ignored; then, the distributed absorption of photons is the only factor of excess noise.

In Appendix A, we derive the spectral characteristics of the photocurrent signal in homodyne detection considering the distributed absorption. The observable of homodyne detection is
\begin{align}
    \hat{I}(\omega) \propto A(\omega)\, \hat{x}_\text{in}(\omega) + B(\omega)\, \hat{x}_\text{vac}(\omega), \label{eq:photocurrent_cont_simp_a}
\end{align}
where
\begin{align}
    A(\omega) &= \left|\int_0^L dx\, e^{-\alpha x} H(\omega; x) \right|, \label{eq:photocurrent_cont_simp_b}\\
    B(\omega) &= \sqrt{\frac{1}{\alpha} \int_0^L \!\!\! dx\, e^{-\alpha x}\Biggl|H(\omega; x) - \alpha e^{\alpha x} \!\!  \int_x^L \!\!\! dx'\, e^{-\alpha x'} H(\omega; x')\Biggr|^2}, \label{eq:photocurrent_cont_simp_c}
\end{align}
and $\hat{I}(\omega)$ is frequency-$\omega$ component of the photocurrent, $\hat{x}(\omega)$ is quadrature operator of sideband field at frequency $\omega$, $H(\omega; x)$ is the Fourier transform of $h(t; x)$, $\hat{x}_\text{vac}(\omega)$ is a vacuum noise term, and $L$ is the thickness of the photodiode. In Eq. (\ref{eq:photocurrent_cont_simp_a}), $[A(\omega)]^2$ is the signal gain spectrum, which shows roll-off due to the transit-time of the carrier drift. $[B(\omega)]^2$ is the gain of vacuum noise term, which express the excess noise originating from the distributed absorption of photons.

Since the power spectrum of shot-noise is given by $[A(\omega)]^2 + [B(\omega)]^2$, the normalization with the shot-noise power gives equivalent optical loss of homodyne detection as
\begin{align}
    \mathcal{L}(\omega) = \frac{[B(\omega)]^2}{[A(\omega)]^2 + [B(\omega)]^2}.
    \label{eq:loss_gain}
\end{align}
If $H(\omega; x)$ is invariant with $x$, $\mathcal{L}(\omega)$ reduces to $e^{-L x}$, expressing the rate of photons that penetrate the intrinsic region. In the limit of $L\rightarrow\infty$, such loss vanishes in $\mathcal{L}(\omega)$. Even in this limit, non-uniform impulse response $H(\omega; x)$ can result in residual of the vacuum term.
Since the excess loss is caused by the variation of the transfer functions respective to the absorption positions, it is non-zero only above the frequency where the transit-time roll-off takes place \cite{PhysRevLett.113.203901}. Note that photodiodes with high quantum efficiency must have enough thickness, and the excess loss term inevitably appears because of the distributed absorption within the absorption distribution $e^{-\alpha x}$, which is characteristic to the photodiode material.

\subsection{Simulation of Excess Loss}
\label{sec:simulation}
The photocurrent impulse response $h(t; x)$ is obtained by the carrier transport dynamics and the Shockley-Ramo theorem. Following assumptions are reasonable when we consider PIN photodiodes with a thin P and N-doped layer, high reverse-bias voltage, low illumination, low frequency, and low I-layer doping condition: (1) Absorption and carrier transport in P and N-doped contact layer can be ignored. (2) Carrier transport is dominated by drift in the depletion region; diffusion, recombination, and scattering contribute little to photocurrent. (3) Electric field $E$ is constant over the I-layer as $E=V/L$ where $V$ is reverse-bias voltage; donor-concentration is small and space-charge effect of photocarriers can be ignored. Then the photocurrent is driven by electrons and holes drifting at constant velocities, leading to a simple formation of $h(t; x)$:
\begin{align}
    \begin{split}
    h(t; x) = &\frac{q}{L} \Biggl[v_\text{p}(E)\,\Theta(t)\,\Theta\left(\frac{x}{v_\text{p}(E)}-t\right)\\
    & \hspace{5em} + v_\text{e}(E)\, \Theta(t)\,\Theta\left(\frac{L-x}{v_\text{e}(E)}-t\right)\Biggr],
    \end{split}
    \label{eq:imp_response_uniformE}
\end{align}
where $q$ is elementary charge, $\Theta(t)$ is the Heaviside step function, $v_\text{e, p}(E)$ is drift velocity of electrons or holes as functions of the electric field.

\begin{figure}[tb]
    \centering
    \includegraphics{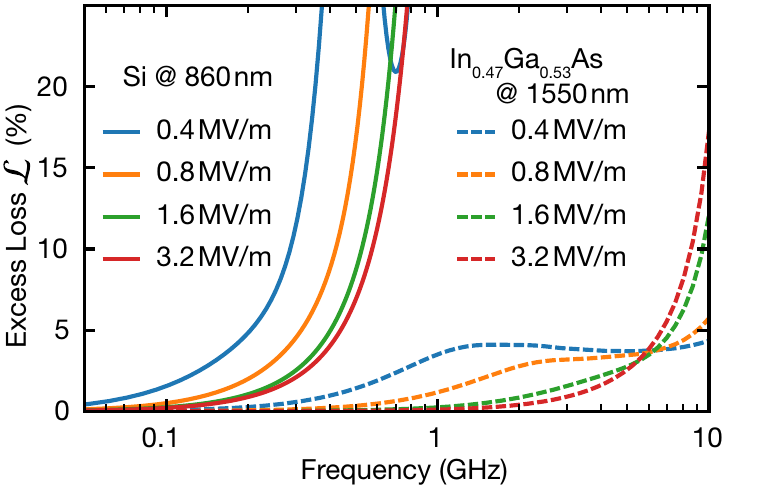}
    \caption{Simulated excess loss spectrum $\mathcal{L}(\omega)$ at different electric fields $E$ of a 100\,\um{} thick Si-PIN photodiode and a 10\,\um{} thick InGaAs-PIN photodiode.}
    \label{fig:loss_spec_sim}
\end{figure}

Based on Eq. (\ref{eq:imp_response_uniformE}), we calculate the excess loss of PIN photodiodes that have near unity quantum efficiency. The excess loss spectrum $\mathcal{L}(\omega)$ of Si with 860\,nm light and that of \ingaas{} with 1550\,nm light is shown in Fig. \ref{fig:loss_spec_sim}. Here we assume vertical incidence of the input light, and ignore the effects of auxiliary layers such as contact layers or buffer layers. In the calculation, following electrical and optical properties of Si or \ingaas{} are used: electronic field dependence of the carrier velocities of Si \cite{sze2001semiconductor} and \ingaas{} \cite{balynas1990time,adachi1992physical}, absorption coefficient of Si at 300\,K \cite{green1995optical} and \ingaas{} at 300\,K \cite{doi:10.1063/1.343580}.

All the traces in Fig. \ref{fig:loss_spec_sim} approach zero at low frequency because the transfer functions $H(\omega; x)$ become independent from $x$. High bias voltage increases the carrier velocity and thereby reduces the excess loss. However, there is a limit due to the saturation of carrier velocity, and thus the minimum loss spectrum is governed by the absorption coefficient and the saturated carrier velocity of the semiconductor material. Si-PIN photodiodes can be considered as loss-less below 100\,MHz and InGaAs-PIN photodiodes work well up to 1\,GHz, thanks to its larger absorption coefficient.

\section{Experiment}
We experimentally verify the above model in the following two ways. We use 860\,nm CW light and a Si-PIN photodiode S5971SPL (Hamamatu Photonics), which has near unity quantum efficiency at 860\,nm and whose active diameter is 0.8\,mm. In both cases, 100\,V reverse-bias voltage is S5971SPL. The saturation characteristics of this photodiodes is separately measured, and we have confirmed that 1\,mW input power induce little saturation below 1\,GHz. The experimental setup is shown in Fig. \ref{fig:expsetup_simple} and the details are in Appendix B.

\begin{figure}[tb]
    \centering
    \includegraphics{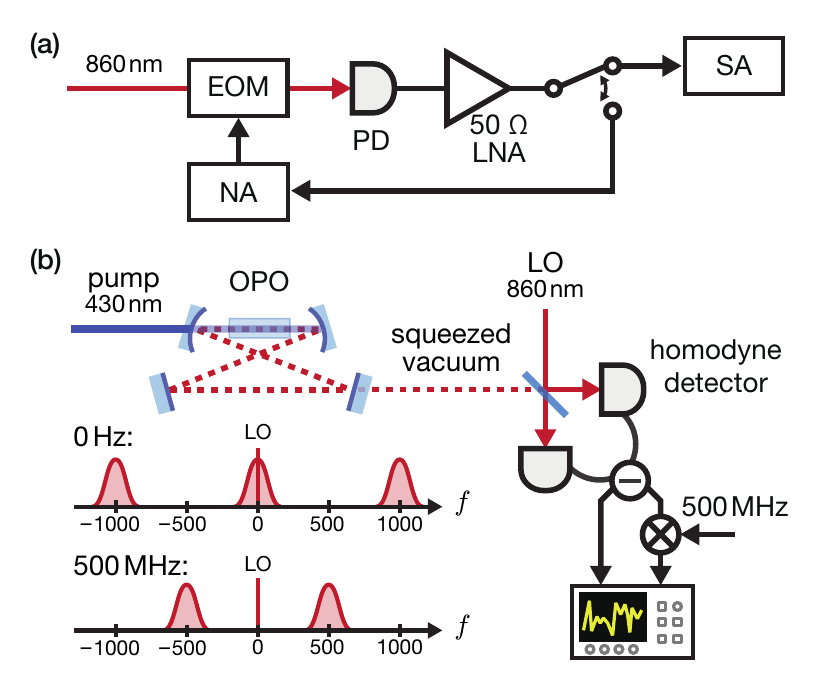}
    \caption{Schematic of the experimental setups. \textbf{(a)} shot-noise and modulation measurement. EOM: electro-optic modulator, NA: network analyzer, SA: spectrum analyzer, LNA: low-noise amplifier. 1\,mW of CW light at 860\, nm is detected by a S5971SPL photodiode. In the measurement of shot-noise, the network analyzer is disconnected and the power spectrum is obtained by the spectrum analyzer. Modulation signal gain is measured by the amplitude modulator and the network analyzer. \textbf{(b)} Squeezed light measurement. OPO: optical parametric oscillator, LO: local oscillator. The OPO is pumped by 430\,nm CW pump and generates a squeezed vacuum at either 0\,Hz and 500\,MHz. The quadrature of squeezed vacuum state is measured by homodyne detection with a carrier frequency LO and the output signal is digitized by an oscilloscope. Resonance structure of the OPO is also shown for the measurement at 0\,Hz and 500\,MHz.}
    \label{fig:expsetup_simple}
\end{figure}

\begin{figure}[tb]
    \centering
    \includegraphics{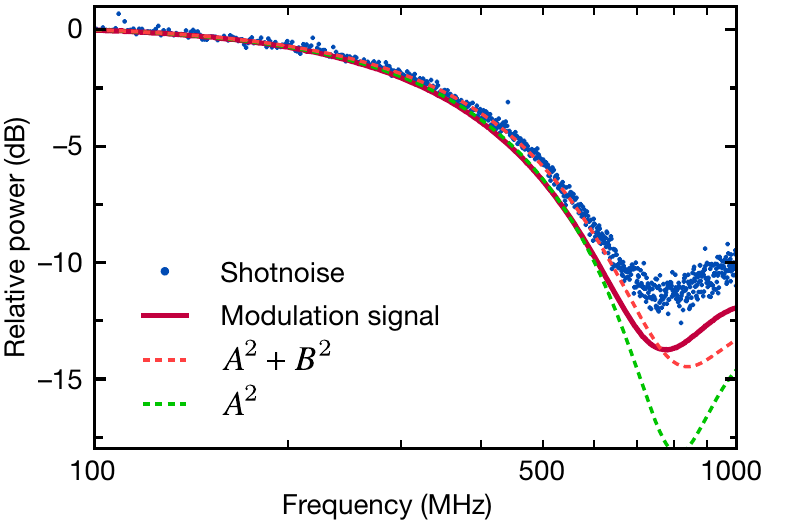}
    \caption{Normalized power spectrum of shot-noise and modulation signal gain of S5971SPL with 1\,mW input power. Theoretical curves are calculated by Eqs. (\ref{eq:photocurrent_cont_simp_a}) and (\ref{eq:imp_response_uniformE}), with the parameters $L=$100\,\um{}, $V=$100\,V.}
    \label{fig:modgain_and_shotgain}
\end{figure}

\textit{(a) Comparison of shot-noise gain with modulation-signal gain of a photodiode}.
Here we consider a one-port homodyne measurement of vacuum state, which comes down to a single-eye detection of a laser beam. In Eq. (\ref{eq:photocurrent_cont_simp_a}), the power spectrum of the input signal $|\hat{x}_\text{in}|^2$ is expressed as $[A(\omega)]^2$ and the shot-noise spectrum is given by $[A(\omega)]^2 + [B(\omega)]^2$. By comparing these spectra, the ratio of $A(\omega)$ and $B(\omega)$ indicates the loss spectrum $\mathcal{L}$ as Eq. (\ref{eq:loss_gain}). These spectra are directly observed by a power spectrum measurement of a single photodiode with continuous light (Fig. \ref{fig:expsetup_simple}a). Here, a very fast photodiode S5973 (Hamamatu Photonics) with 50\,V reverse-bias is used as a reference to calibrate the non-flat gain of the amplifiers, transmission lines, spectrum analyzer, network analyzer, and electro-optic modulator. The S5971SPL's spectrum is divided by the corresponding spectrum of S5973. Figure \ref{fig:modgain_and_shotgain} shows the normalized gain spectrum and theoretical curves. The difference between shot-noise spectrum and signal gain spectrum shows the existence of the excess loss. It is difficult to obtain a precise value of loss, since it is technically challenging to keep high signal-to-noise ratio of the shot-noise over 1\,GHz, and the slight error in power calibration can be considerable in the amount of loss. The simulation curves of $[A(\omega)]^2 + [B(\omega)]^2$ and $[A(\omega)]^2$ well agrees with the experimental data below 600\,MHz. The negative gain peak at 800\,MHz corresponds to a dimple of photocurrent transfer functions $H(\omega; x)$. The mismatch at higher frequency is due to the simplification in Eq. (\ref{eq:imp_response_uniformE}), where each hole and electron current impulse is a rectangular function, corresponding to the transfer function as sinc function. We suppose the assumption, that $E$ is constant over the space-charge region of the photodiode, is not accurate enough to describe the gain characteristics around the peak frequency.

\textit{(b) Measurement of squeezed light at a high-frequency sideband}. A convenient way of measuring the optical loss of photodiodes is to do homodyne measurement of squeezed light \cite{PhysRevLett.117.110801}, because the squeezing level is very sensitive to the absolute value of the detection efficiency. Here we conduct a balanced homodyne measurement of a squeezed vacuum at 500\,MHz, which is generated by an optical parametric oscillator (OPO). The OPO has the free spectral range of 1.0012\,GHz and is locked so that it is resonant at $\pm$500.6\,MHz. A parametric process caused by a CW pump light at 430\,nm produces squeezed light at a 500.6\,MHz double-sideband around the 860\,nm carrier frequency, where in principle a pure squeezed state is obtained \cite{PhysRevA.73.013817}. The purity of the measured squeezed state shows the total amount of losses in the setup. For a high signal-to-noise ratio homodyne measurement of the quadrature at 500.6\,MHz, a resonant-type homodyne detector that equips two S5971SPLs \cite{doi:10.1063/1.5029859} is used. The LO light is 2\,mW of CW light at 860\,nm; thus each S5971SPL has 1\,mW of input power. To calibrate other loss than the PD's high-frequency loss, the squeezing level at 0\,Hz is also measured. Since most of the loss factor such as OPO's internal loss, propagation loss, interference loss at homodyne detection, and imperfect quantum efficiency of the photodiode are common in  0\,Hz and 500\,MHz, we can precisely cancel them by comparing the squeezing levels of these measurements.

\begin{figure}[b]
    \centering
    \includegraphics{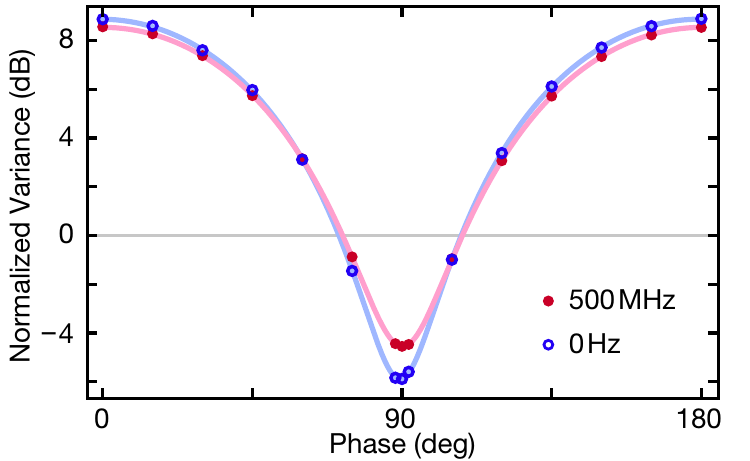}
    \caption{Quadrature variance of the squeezed light at different frequencies. The narrow-band component around the center frequency within 300\,kHz to 1\,MHz is used for the calculation of variance. Statistical uncertainty is smaller than the diameter of points. Fitting curves are calculated by Eq. (\ref{eq:sqlevel}).}
    \label{fig:sqlevel_fit}
\end{figure}

Figure \ref{fig:sqlevel_fit} shows the phase scan plot of the quadrature variance of squeezed light at 500\,MHz and 0Hz. The quadrature variance $V(\theta; \omega)$ normalized by the shot-noise variance is expressed as
\begin{align}
    V(\theta; \omega) = \mathcal{L}(\omega) + \bigl[1-\mathcal{L}(\omega)\bigr]\, \left[\cos^2\theta\, R(\omega) + \frac{\sin^2\theta}{R(\omega)} \right],
    \label{eq:sqlevel}
\end{align}
where $\theta$ is the phase of the quadrature operator and $R(\omega)$ is the initial squeezing level at $\omega$. From the fitting in Fig. \ref{fig:sqlevel_fit}, the optical loss $\mathcal{L}(\omega)$ is estimated at 16.6\% at 0\,Hz and 27.4\% at 500\,MHz, and the initial squeezing level $R(\omega)$ is 9.06\,dB at 0\,Hz and 9.47\,dB at 500\,MHz. The individually measured losses of the setup is summarized in Table \ref{tab:loss}. Note that the total optical loss $\mathcal{L}_\text{tot}$ is calculated from multiple losses $\mathcal{L}_k$ as \(\mathcal{L}_\text{tot} = 1 - \prod_k (1-\mathcal{L}_k)\).

\begin{table}[tb]
  \centering
  \begin{tabular}{lrr}
    \hline\hline
    loss factor                             & 0\,Hz  & 500\,MHz \\
    \hline
    propagation loss                        & 3.5\%  & 3.5\%    \\
    escape efficiency of OPO                & 2.7\%  & 2.8\%    \\
    mode-match at homodyne interferometer   & 6.8\%  & 6.5\%    \\
    gain imbalance in homodyne detection    & 0.0\%  & 0.0\%    \\
    probe light tapping                     & 3.0\%  & 1.7\%    \\
    photodiode quantum efficiency           & 1.7\%  & 1.7\%    \\
    \hline
    total                                   & 16.6\% & 15.3\%   \\
    \hline\hline
  \end{tabular}
  \label{tab:loss}
  \caption{Summary of optical loss in the squeezed light measurement setup. The quantum efficiency of the photodiodes is an estimated value from the squeezed light measurement at 0\,Hz.}
\end{table}

Here, the quantum efficiency of the photodiode S5971SPL is estimated at 98.3\% to make it consistent with the total loss of the squeezed light measurement at 0\,Hz. The difference between the estimated total loss in squeezed light measurement and the individually measured loss is 14.2\% at 500\,MHz, which is to be attributed to the excess optical loss of the photodiode. We assume the about 0.5\% of uncertainty in the estimation of the excess loss considering the fluctuation in the calibration.

\begin{figure}[tb]
    \centering
    \includegraphics{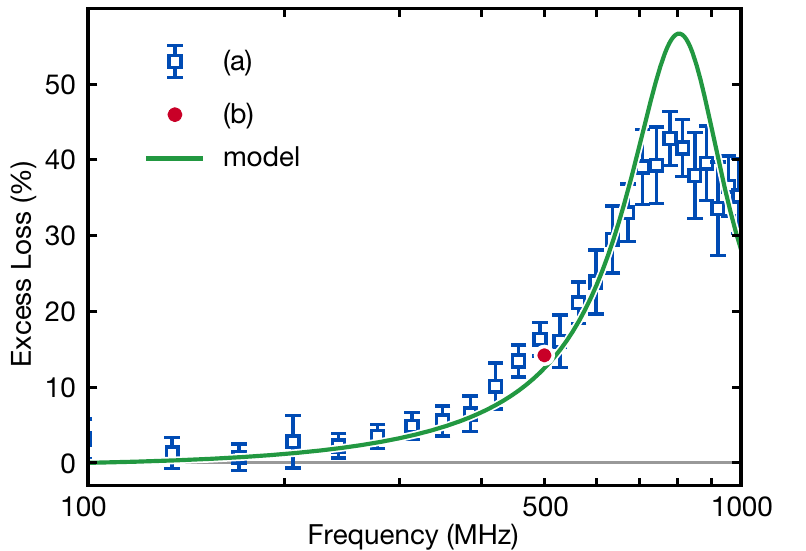}
    \caption{Spectrum of the excess loss of S5971SPL. Blue square points: excess loss calculated from the shot-noise measurement of S5971SPL. Red circle: excess loss estimated from the squeezed light measurement. Green line: 
    theoretical curve from the model in Sec. \ref{sec:simulation}, with the same parameters with Fig. \ref{fig:modgain_and_shotgain}.}
    \label{fig:excesslossspec}
\end{figure}

\textit{(c) Summary of the measurement and simulation of the excess loss of S5971SPL}.
The excess loss spectra of S5971SPL estimated by the two methods are shown in Fig. \ref{fig:excesslossspec}. The simulation based on Sec. \ref{sec:simulation} is also shown. The measured values of the excess loss roughly agree at 500\,MHz. The broadband spectrum of the excess loss shows a characteristic peak around 800\,MHz, where Eq. \ref{eq:imp_response_uniformE} become less reliable around such frequency as we observed in Fig. \ref{fig:modgain_and_shotgain}.

For Si photodiodes at 860\,nm, our model well describes the excess loss, and thus the distributed absorption is dominant in the excess noise. This is because Si has low absorption coefficient, which gives rise to a large variation in the transfer function $H(\omega; x)$ with respect to $x$ within the deep absorption region. Photodiode materials with large absorption coefficient have a smaller effect of distributed absorption and other factors such as carrier scattering may give a contribution to the excess loss. In this sense, Fig. \ref{fig:loss_spec_sim} underestimates the excess loss, especially for InGaAs. A complete prediction of the excess loss can be given by the Monte-Carlo simulation of photocarrier generation and transportation \cite{PhysRevLett.113.203901}.

The 1\,mW input power is small enough to avoid the effect of saturation of the photodiodes. For the high signal-to-noise ratio homodyne detection, higher LO power is preferred; with a high illumination, however, the space-charge effect \cite{williams1994effects} may reduce the bias field and increase the excess loss. This can be easily avoided by broadening the aperture of the photodiodes, though.

In the experiment, the bias field in the photodiode is as large as 1\,MV/m,  and the electron / hole velocity is near to the saturation. There is thus not much room for reducing the excess loss by increasing the reverse-bias (see Fig. \ref{fig:loss_spec_sim}). With a thinner absorption layer, the effect of distributed absorption and carrier transit-time can be suppressed. For example, cavity-enhanced photodiodes \cite{doi:10.1063/1.360322} can realize high quantum efficiency with a very thin absorption layer. However, with the current technology, it is difficult to obtain near 100\% quantum efficiency with such a photodiode structure. Therefore, our PIN photodiode model gives the practical upper limit of the frequency of high-efficiency photodetection.

\section{conclusion}
The distributed photon absorption in photodiodes causes excess noise at high-frequency region of continuous-wave light detection. We have formulated the excess noise by a beamsplitter model of random photo detection, where the noise is re-interpreted as an optical loss. The excess loss at high-frequency region can be a critical limitation in the bandwidth of homodyne detection. We have evaluated the impact of the excess loss for Si and InGaAs photodiodes and shown that it becomes non-negligible above 100\,MHz and 1\,GHz, respectively. We have experimentally verified the excess loss of Si photodiodes by signal gain measurement and squeezed light measurement. These results shows agreement with the theoretical model.

At the 860\,nm band, the material of high-efficiency photodiodes is practically limited to Si and the excess loss is not ignorable above 100\,MHz. For the faster quantum homodyne detection, we would have to choose other wavelength and photodiode material. Another way of increasing the bandwidth of quadrature measurement is to add a parametric amplifier before homodyne measurement to conceal the loss or the noise in the photodetection \cite{PhysRevLett.72.4086,shaked2018lifting}, but it is nonetheless technically challenging to suppress the loss in the parametric amplification.

\section*{Acknowledgements}
This work was partly supported by CREST (No. JPMJCR15N5) of JST, JSPS KAKENHI, and APSA of Japan.

\section*{Appendix A: Formulation of Excess Loss}
\subsection*{Beamsplitter model of distributed photocarrier generation}

\begin{figure}[tb]
    \centering
    \includegraphics{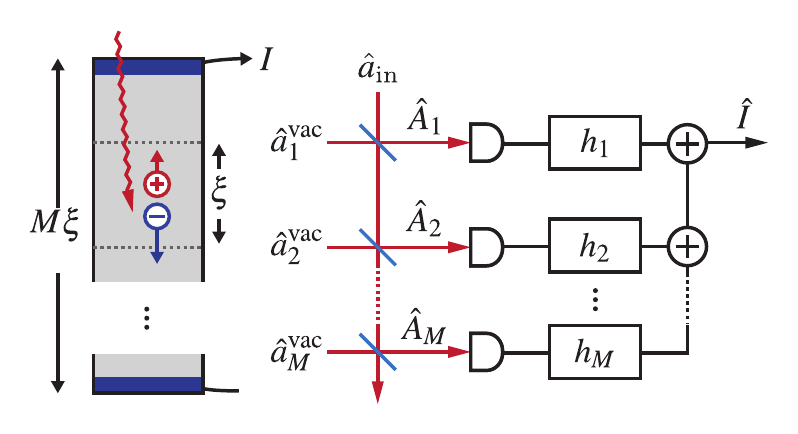}
    \caption{Model of distributed photocarrier generation in the space-charge region of photodiodes. The incident field $\hat{a}_\text{in}$ is probabilistically absorbed in virtually separated thin layers with the thickness $\xi$. The tapped field $\hat{A}_k$ at the $k$-th beamsplitter triggers photocurrent impulses $h_k(t) = h(t; k\xi)$.}
    \label{fig:pd_bs_model}
\end{figure}
By slicing the absorber with the small width $\xi$, photocarrier generation distribution is modeled by a series of beamsplitters with reflectance $r = 1 - e^{-\alpha \xi}$ followed by virtual photon detectors with unity quantum efficiency. Vacuum field $\hat{a}^\text{vac}_k$ irrupts into the virtual detectors from the back ports of the beamsplitters through the transmissivity of $1-r$, representing the randomness of absorption positions.

Other randomness than the distributed absorption such as carrier scattering in the photocarrier transport can increase the excess noise. (Note that recombination is usually included in the evaluation of quantum efficiency). Our model does not include such noises, but they can be also modeled by a set of beamsplitters and impulse responses, which is eventually summarized into the excess loss.

The annihilation operator of the incident field at $k$-th photon detector is written as
\begin{align}
    \begin{split}
        \hat{A}_k(t) = &\sqrt{r}(1-r)^\frac{k-1}{2}\, \hat{a}_\text{in}(t)\\ &-r\sum_{l=1}^{k-1} (1 -r)^\frac{k-l-1}{2}\,\hat{a}^\text{vac}_l(t) + \sqrt{1-r}\,\hat{a}^\text{vac}_{k}(t),
    \end{split} \tag{A1}
    \label{eq:detmodes_d}
\end{align}
where $\hat{a}_\text{in}(t)$ is the annihilation operator of the input field. The propagation delay in the semiconductor layer is ignored here.
The $k$-th photon detector has the photon-to-current impulse response $h(t; k\xi)$ and the output signal is its convolution with the photon number operator $\hat{A}^\dagger_k(t)\hat{A}_k(t)$. The output photocurrent $\hat{I}(t)$ is the sum of each detector's signal, reading
\begin{align}
    \hat{I}(t) = \int d\tau \sum_{k=1}^M \hat{A}^\dagger_k(\tau)\,\hat{A}_k(\tau)\,h(t - \tau; k\xi). \tag{A2}
    \label{eq:pcurrent_d}
\end{align}
The upper bound $M$ of the sum is determined so that $L = M\xi$ is equal to the total thickness $L$ of the intrinsic layer of the photodiode.

\subsection*{Excess loss in homodyne detection}
For optical quadrature measurement, we consider one-port homodyne measurement by introducing large displacement $a_0$ to the incident field as $\hat{a}_\text{in}(t) = a_0 + \hat{a}_\text{sig}(t)$, where $\hat{a}_\text{sig}(t)$ is a small signal field operator. Balanced homodyne measurement is straightforwardly derived from one-port homodyne measurement, given that the local-oscillator light and photocurrent response is well balanced.

Taking the first-order term of $a_0$ in Eq. (\ref{eq:pcurrent_d}) and ignoring the DC offset, we have
\begin{align}
    \frac{\hat{I}(t)}{a_0} = \int d\tau \sum_{k=1}^M \sqrt{r(1-r)^k}\,\bigl[\hat{a}_k^\dagger(\tau) + \hat{a}_k(\tau)\bigr] h(t - \tau; k\xi). \tag{A3}
    \label{eq:selfhom_current}
\end{align}
The AC component of the quadrature signal is given by rock-in detection of $\hat{I}(t)$. Here we focus on in-phase component of the homodyne signal $\hat{I}_\text{I}(\omega) \equiv \bigl[\hat{I}(\omega) + \hat{I}(-\omega)\bigr]/2$.
Fourier transform of Eq. (\ref{eq:selfhom_current}) yields
\begin{align}
    \frac{\hat{I}(\omega)}{a_0} = \sum_{k=1}^M \sqrt{r(1-r)^k}\,\bigl[\hat{a}_k^\dagger(\omega) + \hat{a}_k(\omega)\bigr] H(\omega; k\xi), \tag{A4}
    \label{eq:selfhom_current_freq}
\end{align}
where $\hat{a}(\omega)$ is frequency-mode annihilation operators and $H(\omega; x)$ is the position-dependent transfer function. Then the in-phase photocurrent expresses the quadrature amplitude of a double-sideband mode:
\begin{align}
    \begin{split}
        \frac{\hat{I}_\text{I}(\omega)}{a_0} =\ 
        & r\left|\sum_{k=1}^M(1-r)^{k-1}\,H(\omega; k\xi) \right|\hat{x}_\text{in}^{\theta_\omega}(\omega) \\
        & - \sqrt{r} \sum_{k=1}^M\, \Biggl| (1-r)^\frac{k}{2} H(\omega; k\xi) \\
        &\hspace{4em} - r\sum_{l=k}^M(1-r)^{1-\frac{k}{2}} H(\omega; l\xi)\Biggr|\, \hat{x}_k(\omega),
    \end{split} \tag{A5}\\
    \theta_\omega =\ &\arg \left[\sum_{k=1}^M(1-r)^{k-1}\,H(\omega; k\xi)\right], \tag{A6}
\end{align}
where $\hat{x}_\text{in}^\theta(\omega) \equiv \bigl[e^{-i\theta}\hat{a}^\dagger_\text{in}(-\omega) + e^{i\theta}\hat{a}_\text{in}(\omega)\bigr]/2$ is quadrature operator of sideband field at $\omega$, and $\hat{x}_k^\theta \equiv \bigl[e^{-i\theta}\hat{a}^\dagger_k(-\omega) + e^{i\theta}\hat{a}_k(\omega)\bigr]/2$ is quadrature operators of the back port vacuum field, which is invariant with $\theta$. By taking the limit of $\xi \rightarrow 0$, the above equations become
\begin{align}
    \frac{\hat{I}_\text{I}(\omega)}{a_0} &= A(\omega)\,\hat{x}^{\theta_\omega}_\text{in}(\omega) + B(\omega)\, \hat{x}_\text{vac}(\omega), \label{eq:photocurrent_cont_a} \tag{A7}\\
    \theta_\omega &= \arg \int_0^L dx\, e^{-\alpha x} H(\omega; x), \tag{A8}
\end{align}
where $A(\omega), B(\omega)$ is given by Eqs. (\ref{eq:photocurrent_cont_simp_b}), (\ref{eq:photocurrent_cont_simp_c}), and the vacuum noise terms are represented by a single vacuum operator $\hat{x}_\text{vac}(\omega)$ using the relationship
\begin{align}
    \lim_{\xi\rightarrow 0} \sqrt{\xi} \sum_k f(k\xi)\, \hat{x}_k(\omega) = \sqrt{\int dx\, \bigl|f(x)\bigr|^2}\, \hat{x}_\text{vac}(\omega) \tag{A9}
\end{align}
for vacuum operators $\hat{x}_k(\omega)$ and a real function $f(x)$. By omitting explicit notation of sideband phase $\theta_\omega$, Eq. (\ref{eq:photocurrent_cont_a}) reduces to Eq. (\ref{eq:photocurrent_cont_simp_a}). The quadrature-phase component $\hat{I}_\text{Q}(\omega)$ of the sideband signal is similarly derived as
\begin{align}
    \frac{\hat{I}_\text{Q}(\omega)}{a_0} &= A(\omega)\,\hat{x}^{\theta_\omega+\pi/2}_\text{in}(\omega) + B(\omega)\, \hat{x}'_\text{vac}(\omega), \tag{A10}
\end{align}
where $\hat{x}'_\text{vac}(\omega)$ is another vacuum operator that is independent of $\hat{x}_\text{vac}$.

\section*{Appendix B: Details of the Experiment}
\subsection{Comparison of shot-noise gain with intensity modulation signal gain of a photodiode}
\begin{figure}[tb]
    \centering
    \includegraphics{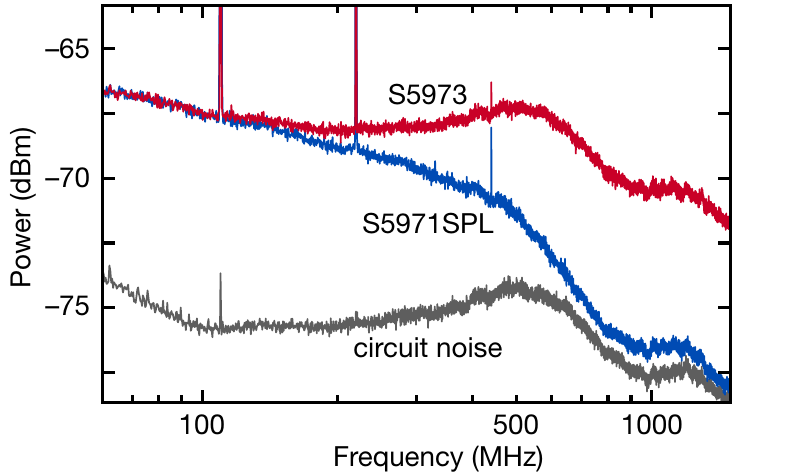}
    \caption{shot-noise spectrum detected by S5971SPL and S5973. The circuit noise of the detector is also shown.}
    \label{fig:rawspectrum}
\end{figure}

CW laser light at 860\,nm is produced by a Ti:Sapphire laser MBR-110 (Coherent). It is filtered by two mode-cleaning cavities to reduce the intensity noise above 10\,MHz down to the shot-noise level. The photodetector is composed of a series of 50\,$\Omega$-input low-noise amplifiers HMC8410 (Analog Devices), which has 1.1\,dB of noise figure, and ZFL-1000LN+ (Minicircuits). The power spectrum of the shotonise signal is obtained by a spectrum analyzer E4402B (Agilent Technologies) with the resolution bandwidth of 300\,kHz and the average count of 1000. The raw trace data of the shot-noise spectra is shown in Fig. \ref{fig:rawspectrum}. We just replaced the photodiode from S5971SPL to S5973 on the detector board to take the two traces with the same electric gain. In Fig. \ref{fig:modgain_and_shotgain} the circuit noise is subtracted from the shot-noise power. The signal-to-noise ratio of the shot-noise signal of S5971SPL is smaller than 5\,dB above 500\,MHz; this is why the uncertainty of the noise-subtracted shot-noise power gets large at high-frequency. The sharp peaks in the traces are intensity noise of 110\,MHz modulation and its harmonics in the laser light. We eliminated these peaks from the data for the statistical processing for Fig. \ref{fig:excesslossspec}. The modulation signal gain is measured by a network analyzer E5080A (Keysight Technologies) and a waveguide intensity modulator (EOSPACE). 

\subsection{Measurement of squeezed light at high-frequency sideband}
The OPO is a bow-tie type cavity with a type-0 phase-matched PPKTP crystal (Raicol Crystals). The reflectivity of the output coupler is 88\% and the linewidth of the OPO cavity is 10\,MHz.
The OPO is locked by a Pound-Drever-Hall method with a counter-propagating locking beam, which is detuned at a certain frequency to produce squeezed light at the specific frequencies. The oscillation threshold of the OPO is measured at 550\,mW and the pump power is set at 100\,mW, corresponding to the normalized pump amplitude 0.43. The relative phase between the pump and the LO is locked during the acquisition. The measurement setups of the 0\,Hz squeezed light measurement and the 500\,MHz squeezed light measurement are slightly different and they are separately described below.

The output signal of the homodyne detection is digitized by a 12-bit oscilloscope DSOS204A (Keysight Technologies) with the sampling rate of 200\,MHz and duration of 150\,\us{}.
The power spectra are obtained by the fast Fourier transform and is averaged over 1000 frames of the traces for each LO phases. The power spectra of the squeezed light are shown in Fig. \ref{fig:rawspectrum}. In Fig. \ref{fig:sqlevel_fit}, the power averages of the spectrum within 300\,kHz-1\,MHz are shown, where the circuit noise power is subtracted from the signals in frequency-domain.

\begin{figure}[tb]
    \centering
    \includegraphics{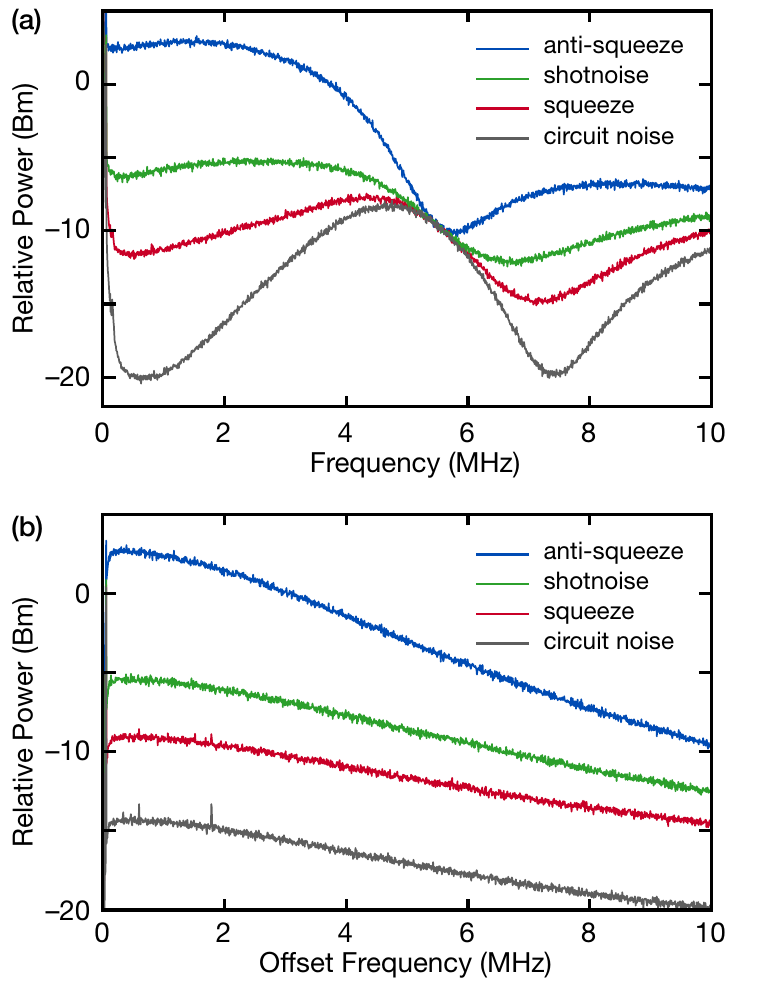}
    \caption{Power spectrum of squeezed / anti-squeezed light at \textbf{(a)} 0\,Hz and \textbf{(b)} 500\,MHz. The shot-noise level and circuit noise level are also shown.}
\end{figure}

\textit{(a) Squeezed light measurement at 500\,MHz}.
The OPO-locking beam is detuned at 1.5018\,GHz. To probe the pump phase, a reference light with a modulation at $\pm500.6$\,MHz generated by an EOM is introduced from the high-reflection port of the OPO. The reference modulation component is downconverted to 0\,Hz by an EOM placed on the output beam. A triangle-shaped frequency separator cavity extracts the 0\,Hz component, which is subsequently detected for the feedback control of the pump phase. The modulation cause 1.7\% of optical loss in the in-phase component of the squeezed light at 500.6\,MHz, which is took into account in Table. \ref{tab:loss}. The gain and phase balance of the homodyne detection is precisely adjusted and we obtained 40\,dB of intensity signal cancelation in the LO. Here the arm length of beams are well matched and the gain of the photodiodes at 500\,MHz are slightly tuned by adjusting the reverse voltage: 102.5\,V or 97.5\,V are applied at each S5971SPL. The signal-to-noise ratio of the shot-noise against the circuit noise is roughly 9\,dB over the measured frequency range. The output of the homodyne detector is processed by an IQ demodulator ADL5380 (Analog Devices) with a 500.6\,MHz electric LO to obtain in-phase component of the sideband signal. Note that the power transfer of the reference light downconversion depends on the sideband phase: the quadrature phase component of 500.6\,MHz sideband signal undergoes one-third of the loss of the in-phase component. We avoid the mixing of these phase components by choosing the in-phase component by matching the demodulation phase with the downconversion phase.

\textit{(b) Squeezed light measurement at 0\,Hz}.
The OPO-locking light is detuned at 1.012\,GHz. A reference light is introduced at 0\,Hz from the high-reflection port of the OPO. The reference light is tapped by a partial-reflection mirror with 97\% reflectivity in place of the frequency separator cavity, which causes 3.0\% of the optical loss on the squeezed light. All the other optics are common with the 500\,MHz squeezed light measurement. The EOMs are electrically turned off. The DC output of the homodyne detector is directly digitized by the oscilloscope with the same configuration. The DC signal is affected by the resonant circuit in the homodyne detector and the gain spectrum has a unique shape as shown in Fig. \ref{fig:rawspectrum}.


%

\end{document}